\documentclass[aps,prc,superscriptaddress,nofootinbib,floatfix,preprint]{revtex4}

\usepackage{graphicx}	

\begin{document}

\title{Exploring the Proton's Spin at PHENIX}

\author{C. Aidala for the PHENIX Collaboration}
\affiliation{Columbia University\\ 
New York, New York 10027, USA\\
E-mail: caidala@bnl.gov}

\date{October 21, 2004}

\begin{abstract}
In late 2001 the first polarized proton collisions at the Relativistic Heavy Ion Collider (RHIC) took place.  
The PHENIX experiment at RHIC has a broad program to investigate the spin structure of the proton.  This program 
will be described, and first results will be presented.
\end{abstract}

\maketitle


\section{Introduction}

Far from the point particle it was once believed to be, the proton has
proven to be an extremely complex entity. \ A very rich structure has
gradually been uncovered over the past 40 years of research. \ A thorough
comprehension of proton structure, in particular its spin structure, remains
the goal of extensive ongoing study. \ The PHENIX experiment at the
Relativistic Heavy Ion Collider (RHIC) at Brookhaven National Laboratory is
in a position to make significant contributions to further understanding the
origin of the proton's spin. \ 

\section{History of Proton Structure}

\subsection{The Quark-Parton Model}

In the 1960s, in deep-inelastic scattering (DIS) experiments at SLAC
analogous to the famous Rutherford scattering experiment that led to the
discovery of the atom's hard core, it was found that protons also had "hard"
subcomponents \cite{Bloom:1969kc,Breidenbach:1969kd}. \ These hard subcomponents came to be
known as partons. \ It took some time before the experimentally observed
partons inside the proton came to be identified as the theoretically
hypothesized quarks, but eventually the quark-parton model of the proton
came into being. \ As experimental work progressed and higher-energy lepton
beams were used as probes, the proton came to reveal a much more intricate
structure than that of the three so-called "valence" quarks. \ These other
subcomponents are now known to be sea quarks and gluons. \ 

\subsection{The Spin Structure of the Proton}

For many years it was assumed that the proton's spin of $\frac{1}{2}\hbar $
was due to the spins of the three spin-$\frac{1}{2}$ valence quarks, with
two oriented in one direction and one in the other. \ In the late 1980s,
however, the EMC experiment at CERN \cite{Ashman:1987hv}\ discovered that only $12\pm
16\%$ of the proton's spin was carried by quarks. \ This surprising result
became known as the "proton spin crisis". \ Subsequent experimental work,
mostly through DIS, has continued to explore this problem for more than 25
years, yet there remains much to be understood. \ In particular, the
magnitude and even sign of the gluon spin's contribution to the spin of the
proton remains to be determined, the flavor breakdown of the sea quarks'
contributions is largely unknown, and the contribution from orbital angular
momentum of both quarks and gluons has yet to be probed. \ 

\section{The Relativistic Heavy Ion Collider (RHIC)}

\subsection{RHIC Physics}

RHIC is the most versatile hadronic collider in the world. \ It is capable
of colliding heavy ions up to $\sqrt{s}$ = 200 GeV/nucleon and polarized
protons up to $\sqrt{s}$ = 500 GeV, as well as asymmetric species. \
Collision of asymmetric species is possible due to independent rings with
independent steering magnets. \ In the first four years of running, RHIC has
provided gold collisions at four different energies, deuteron-gold
collisions, and polarized proton-proton collisions, with plans for copper
collisions in the upcoming year. \ The flexibility of RHIC allows for an
extremely diverse physics program. \ The heavy-ion physics program
investigates strongly-interacting matter under extreme conditions of density
and temperature. \ Systematic variations of nuclear matter with collision
species and energy are being examined, and nucleon structure in a nuclear
environment is being studied. \ 

The polarized proton program seeks a better understanding of the proton's
spin structure, in particular contributions from the gluons and sea quarks.
\ By studying hadronic collisions rather than deep-inleastic scattering,
RHIC experiments may directly observe gluon-scattering processes. \ As a
collider, RHIC can provide collisions at much higher energy than can be
achieved in fixed-target measurements. \ As a result hard processes,
describable by perturbative QCD (pQCD), can be studied, and new probes such
as W bosons will eventually become available. The application of
factorization to pQCD processes is of particular importance because\ it
allows one to separate out parton distribution functions (pdf's), partonic
hard-scattering cross sections, and fragmentation functions (FF's). \
Partonic hard-scattering cross sections are directly calculable in pQCD,
while pdf's and FF's must be determined experimentally. \ RHIC experiments
have access to a wealth of data from other experiments on pdf's and FF's,
allowing them to check the applicability of pQCD calculations to their
unpolarized data and subsequently utilize factorized pQCD to determine
various polarized pdf's more accurately.

\begin{figure}[thb]
 \includegraphics[width=0.55\linewidth]{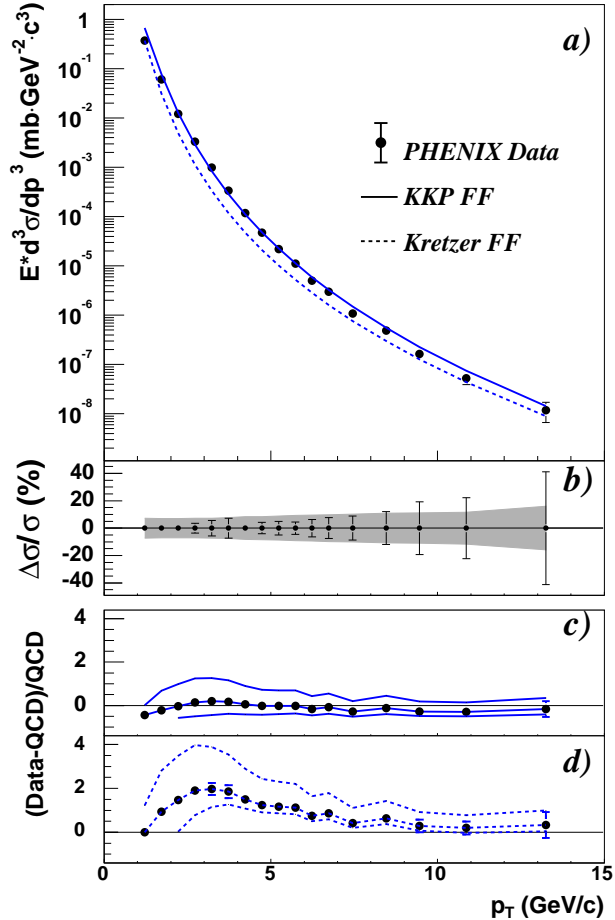}  
 \caption{\label{fig:pi0cross_nlo}PHENIX results (points) for the
invariant differential cross section for inclusive neutral pion production
at 200 GeV. \ In panel (b) the relative statistical (points) and
point-to-point systematic (band) errors are shown. The curves are the results
from NLO pQCD calculations using two different sets of fragmentation
functions. \ See \cite{Adler:2003pb} for more details.}
 \end{figure}

\begin{figure}[thb]
 \includegraphics[width=0.65\linewidth]{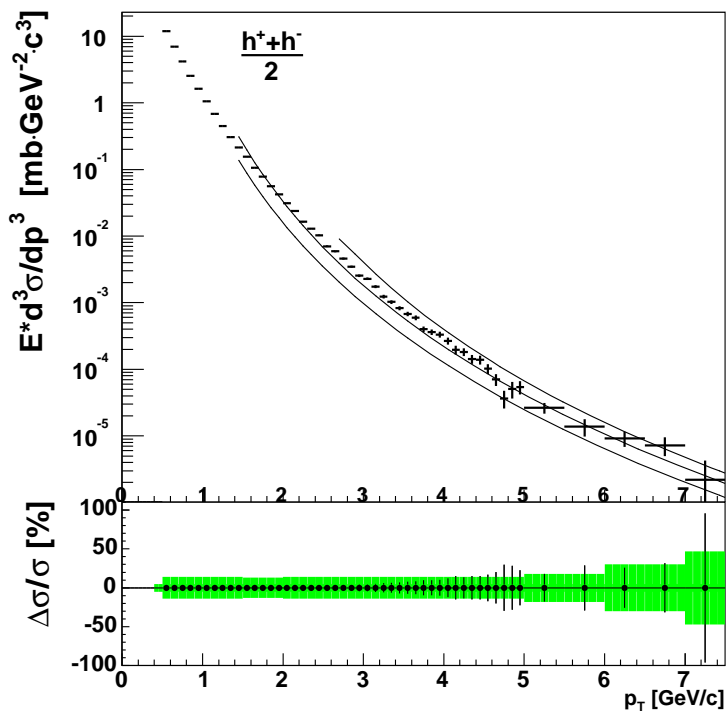}  
 \caption{\label{fig:chargedcross}Preliminary PHENIX results
(points) for the invariant differential cross section for inclusive charged
hadron production at 200 GeV. \ In the bottom panel the relative statistical
(points) and point-to-point systematic (band) errors are shown. \ The curves
indicate NLO pQCD calculations by W. Vogelsang, using renormalization scales
of $\frac{p_{T}}{2}$, $p_{T}$, and $2p_{T}$.}
 \end{figure}

The unpolarized cross sections for mid-rapidity production (see Figure 
\ref{fig:pi0cross_nlo}) as well as forward production of neutral pions have been
measured in 200-GeV proton-proton collisions at RHIC and have been found to
agree well with next-to-leading order (NLO) pQCD calculations 
\cite{Adler:2003pb, Adams:2003fx}. In addition, as shown in Figure 
\ref{fig:chargedcross}, there are preliminary results from PHENIX for
mid-rapidity production of inclusive charged hadrons which also demonstrate
consistency with NLO pQCQ. \ This agreement indicates that NLO pQCD will be
applicable in interpreting polarized data from RHIC as well and provides a
solid theoretical foundation for the spin physics program.

\subsection{RHIC as a Polarized Proton Collider}

RHIC is the first high-energy polarized proton collider in the world. \ This
achievement is possible due to the development of a variety of technologies
to create, maintain, and measure the beam polarization throughout
acceleration and storage.

\paragraph{RHIC-AGS complex}

For proton-proton running, the path traveled by the protons is through a
linac, a booster, the Alternating Gradient Synchrotron (AGS), and finally
RHIC. \ The polarized source reliably provides a polarization of
approximately 80\%. \ The polarization in the AGS is maintained via careful
tuning to avoid depolarizing resonances during acceleration and a partial
Siberian snake. \ Siberian snakes, helical magnets developed at Novosibirsk,
rotate the spin vector of the proton $180^{\circ }$\ such that any effects
of depolarizing resonances will effectively be averaged out on the next turn
around the ring. \ In 2005 a full-length, superconducting Siberian snake
will be installed, completing the array of equipment related to running
polarized protons in RHIC. \ Once the superconducting snake is installed,
RHIC should be capable of reaching its design beam polarization of 70\% at a
beam energy of 250 GeV. \ In RHIC, there are two Siberian snakes installed
in each ring. \ Very little polarization loss has been observed in the RHIC
rings through acceleration and storage.

\paragraph{Polarimetry}

The RHIC-AGS complex utilizes various polarimeters to determine the beam
polarization at different points along its path. \ In particular, there are
proton-carbon (pC) polarimeters in both the AGS and RHIC which make use of
Coulomb nuclear interference (CNI) \cite{Jinnouchi:2003cp}. \ A filament of carbon is
inserted into the proton beam, and the left-right (azimuthal) asymmetry of
recoil carbon atoms from $p^{\uparrow }C\rightarrow p^{\uparrow }C$ elastic
scattering is measured. \ The analyzing power, $A_{N}\approx 0.015$,
originating from the anomalous magnetic moment of the proton is exploited. \
The polarization of the beam can be determined from the following set of
equations, in which $N_{L}$ ($N_{R}$) is the number of recoil carbon atoms
observed to the left (right) of the beam :

\[
P_{Beam}=\frac{\varepsilon _{LR}}{A_{N}},\varepsilon _{LR}=\frac{N_{L}-N_{R}%
}{N_{L}+N_{R}} 
\]

The uncertainty on the polarization measurement from the CNI polarimeters is
currently approximately 30\%, and the analyzing power must be further
calibrated for improved polarization measurements.

In the spring of 2004, a hydrogen-jet polarimeter \cite{Zelenski:2003dn}\ was
commissioned. \ A polarized hydrogen gas jet target is inserted into the
beam, and the left-right asymmetry in p-p elastic scattering is measured in
the CNI regime. \ The hydrogen-jet polarimeter will be used to calibrate the
pC polarimeters and is expected to reduce the uncertainty on the
polarization from \symbol{126}30\% to 5\%.

\paragraph{Spin direction}

The naturally stable spin direction is transverse to the proton's momentum,
in the vertical direction. \ Spin-rotator magnets immediately outside the
STAR and PHENIX interaction regions are used to achieve longitudinal spin. \
These magnets were not commissioned until 2003, so during the 2001-2 run
only data with transverse spin were taken. \ A detector which exploits
previously measured forward-neutron azimuthal asymmetries in transverse-spin
collisions is used to confirm the longitudinal component of the spin at the
PHENIX interaction region.

\subsection{The PHENIX Experiment}

There are four major experiments at RHIC: two larger experiments, PHENIX and
STAR, and two smaller ones, BRAHMS and PHOBOS. \ PHENIX, STAR, and BRAHMS
all have spin physics programs. \ The four experiments have capabilities
that overlap in many areas, making it possible to corroborate new results,
but also areas of specialization which make the experiments complementary.

The PHENIX collaboration is comprised of approximately 480 participants from
12 different nations. \ The PHENIX detector \cite{Adcox:2003zm} consists of
two central spectrometer arms to track charged particles and detect
electromagnetic processes, two forward spectrometer arms to identify and
track muons, and three global detectors to determine when a collision
occurs. \ The central arms cover a pseudorapidity range of $|\eta |<0.35$
and $90^{\circ }$ in azimuth each, while the forward arms cover $1.2<|\eta |<%
\symbol{126}2.2$ and $2\pi $ in azimuth. \ PHENIX was specifically designed
to have a high rate capability and high granularity as well as good mass
resolution and particle-identification capabilities.

\subsection{The Spin Physics Program at PHENIX}

The first polarized proton collisions at RHIC were achieved in December
2001. \ In the 2001-2002 run, an average beam polarization of 15\% was
achieved, and 150 $nb^{-1}$ of transverse-spin data were collected by
PHENIX. \ In 2003, the average polarization reached 27\%, and 220 $nb^{-1}$
of longitudinal-spin data were taken. \ 2004 was principally a commissioning
run to improve the polarization in the AGS and to commission the hydrogen
jet polarimeter. \ During four days of data taking at the end of the
commissioning period 75 $nb^{-1}$ with an average polarization of
approximately 40\% were collected. \ There has been tremendous progress in
machine performance over the first three years of running polarized protons
at RHIC, and an extensive spin run of approximately 10 weeks with close to
50\% polarization is anticipated in 2005. \ Proton running up until this
point has been at $\sqrt{s}=200$ GeV; \ 500-GeV runs are planned for the
future.

PHENIX has a broad spin physics program. \ The principal areas of
investigation are the gluon polarization ($\Delta G$), flavor separation of
the sea quark polarization ($\Delta \overline{u}$, $\Delta \overline{d}$),
and transverse spin physics. \ PHENIX will be able to access a number of
channels which probe $\Delta G$ through double longitudinal-spin
asymmetries. \ These channels include pion production, for which results
have already been published (see below), prompt photon production, dominated
by gluon Compton scattering, heavy flavor production, mainly from
gluon-gluon fusion, and jet production. \ When RHIC begins running 500-GeV
protons, PHENIX will have access to W bosons, which will be identified via
their leptonic decay mode. \ Because W$^{+}$ (W$^{-}$) production will be
almost entirely from $u+\overline{d}$ ($\overline{u}+d$) and $\Delta u$ and $%
\Delta d$ are already well known, it will be possible to single out $\Delta 
\overline{u}$ and $\Delta \overline{d}$ from measurement of the single
longitudinal-spin asymmetry of W production. \ The transverse spin physics
program at PHENIX seeks to understand the transverse spin structure of the
proton. \ This structure will be explored through a variety of measurements,
including single transverse-spin asymmetries, for which there are already
results (see below), jet correlations, the double transverse-spin asymmetry
of the Drell-Yan process, and the interference fragmentation of pion pairs.

\subsection{Recent Spin Physics Results}

From the first two years of polarized proton collisions at RHIC, PHENIX has
results on the single transverse-spin asymmetry of neutral pions and charged
hadrons as well as the double longitudinal-spin asymmetry of neutral pions 
\cite{Aidala:2004du}, \cite{Adler:2004ps}. \ 

\subsubsection{Single Transverse-Spin Asymmetry of Neutral Pions and Charged
Hadrons}

The single transverse-spin asymmetry in the yield of a particular particle
is given by 
\[
A_{N}=\frac{1}{P_{beam}}\left( \frac{N_{L}-N_{R}}{N_{L}+N_{R}}\right) 
\]%
where $N_{L}$ ($N_{R}$) is the particle yield to the left (right) of the
polarized beam.

Large single transverse-spin asymmetries on the order of 20-30\% have been
observed in a number of experiments \cite{Adams:1991cs,Airapetian:1999tv,Adams:2003fx}, ranging in
energy from $\sqrt{s}$~=~20-200~GeV. The large asymmetries seen have
stimulated more careful study by the theoretical community of polarized
cross sections, in particular their dependence on the intrinsic transverse
momentum of the partons ($k_{T}$) (see e.g. \cite{Mulders:1995dh}).

Over the years, a number of models based on pQCD have been developed to
predict these $k_{T}$ dependencies and to explain the observed asymmetries.
Among these models are the Sivers effect \cite{Sivers:1989cc,Sivers:1990fh}, transversity and the
Collins effect \cite{Collins:1992kk}, and various models which attribute the
observed asymmetries to higher-twist contributions (see e.g. \cite{Qiu:1998ia}). \
The Sivers effect hypothesizes that the asymmetries are due to
spin-dependent intrinsic partonic momentum; \ the Collins effect suggests
that they stem from a spin-dependent transverse momentum kick in the
fragmentation process. \ The Collins effect requires transversity, the
degree to which quarks in a transversely polarized proton are transversely
polarized, to be non-zero in order to produce a non-zero asymmetry.

\begin{figure}[thb]
 \includegraphics[width=0.65\linewidth]{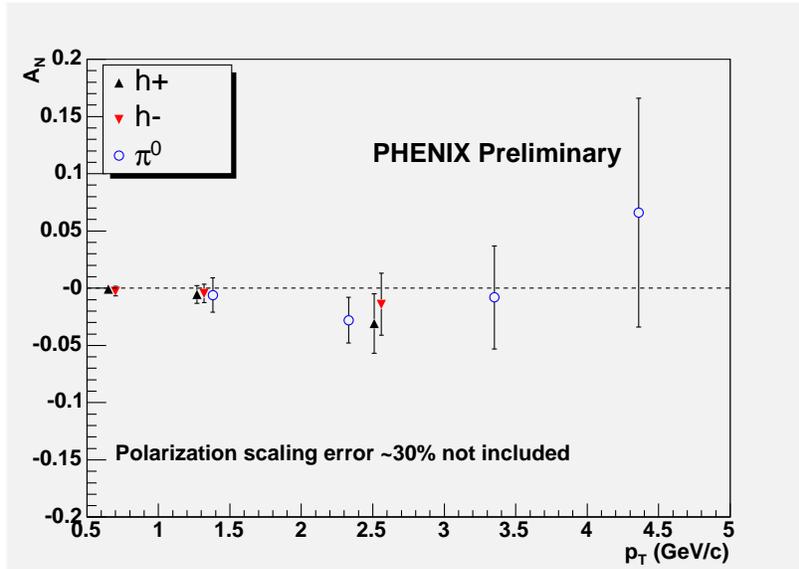}  
 \caption{\label{fig:singleAsym}Preliminary PHENIX results as a
function of transverse momentum for the single transverse-spin asymmetry of
inclusive charged hadrons and neutral pions at mid-rapidity ($x_{F}\approx 0$%
). \ The positive charged hadron points are shifted 50 MeV/c to the left for
readability. \ See \cite{Aidala:2004du} for more details.}
\end{figure}

In Figure \ref{fig:singleAsym} preliminary PHENIX data on the transverse
single-spin asymmetry of inclusive charged hadrons as well as neutral pions
is shown as a function of $p_{T}$. \ The data were taken with the central
arms and thus represent a pseudorapidity coverage of $|\eta |<0.35$,
corresponding to $x_{F}=\frac{p_{L}}{\sqrt{s}/2}\approx 0$. \ The single
transverse-spin asymmetries observed for production of both neutral pions
and inclusive charged hadrons at $x_{F}\approx 0$ are consistent with zero
over the measured transverse momentum range. \ A small asymmetry in this
kinematic region follows the trend of previous results, which indicate a
decreasing asymmetry with decreasing $x_{F}$ \cite{Adams:1991cs,Adams:1994yu,Adams:2003fx}. \
As a significant fraction of neutral pion production in this kinematic
region comes from gluon scattering, any contribution to the asymmetry from
transversity and the Collins effect, requiring a scattered quark, would be
suppressed, while contributions from the Sivers effect or other mechanisms
would remain a possibility. \ Further theoretical study of the results will
have to be performed in order to interpret their full implications for the
transverse spin structure of the proton. \ Future measurements reaching
higher $p_{T}$ will be dominated instead by quark scattering and a better
probe of transversity and the Collins effect. \ See \cite{Aidala:2004du} for
further discussion of these results.

\subsubsection{Double Longitudinal-Spin Asymmetry of Neutral Pions}

\begin{figure}[thb]
 \includegraphics[width=0.65\linewidth]{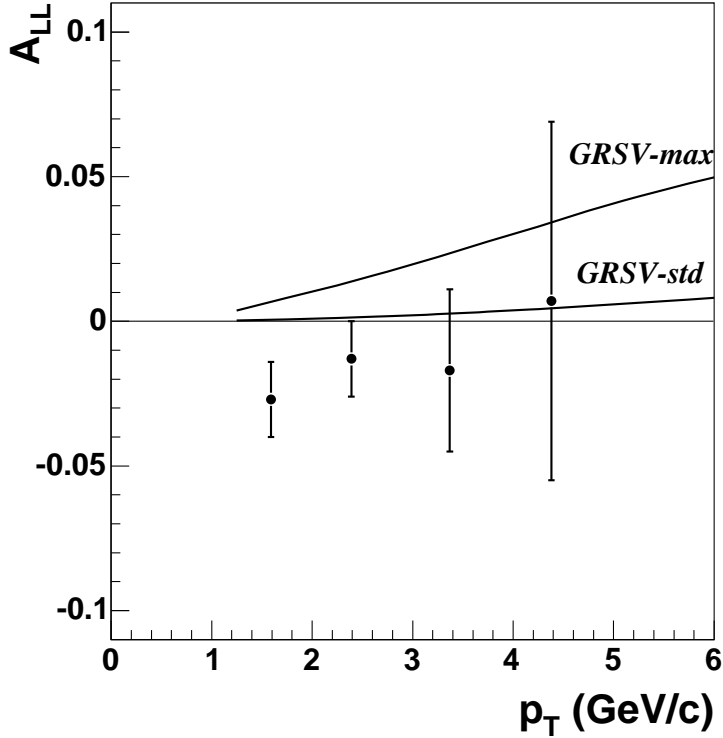}  
 \caption{\label{fig:doubleAsym}PHENIX results for the double
longitudinal-spin asymmetry of inclusive neutral pions at mid-rapidity. \ A
scale uncertainty of $\pm 65\%$ is not included. \ Two theoretical
calculations based on NLO pQCD are shown for comparison with the data. \ See 
\cite{Adler:2004ps} for more details.}
\end{figure}

From the 2003 data-taking period, PHENIX obtained its first results probing $%
\Delta G$, the gluon spin contribution to the spin of the proton. \ The
double longitudinal-spin asymmetry of neutral pions was measured at
mid-rapidity. \ The double longitudinal-spin asymmetry is given by

\[
A_{LL}=\frac{1}{|\langle P_{1}P_{2}\rangle |}\left( \frac{N_{++}-RN_{+-}}{%
N_{++}+RN_{+-}}\right) 
\]%
where $P_{1}$ and $P_{2}$ are the beam polarizations, $N_{++}$ ($N_{+-}$) is
the particle yield from same-helicity (opposite-helicity) bunch crossings
and $R$ is the relative luminosity between same- and opposite-helicity
crossings. \ In Figure \ref{fig:doubleAsym}, the double-longitudinal
asymmetry of neutral pions is shown as a function of $p_{T}$. \ The curves
indicate two theoretical calculations based on NLO pQCD. \ The data points
do not suggest a large contribution from gluon spin. \ For further details
regarding the analysis and these results, see \cite{Adler:2004ps}.

As mentioned above, in the current kinematic range $\pi ^{0}$ production has
a significant contribution from g-g scattering. \ This gluon dominance makes
the $A_{LL}^{\pi ^{0}}$ measurement quite sensitive to the polarized gluon
pdf; \ however, because the polarized gluon pdf enters the factorized cross
section twice at approximately equal values of $x_{Bj}$, it is not
straightforward to determine the sign of $\Delta G$ from this measurement. \
\ Further theoretical discussion of these results and the sign of $\Delta G$
can be found in \cite{Jager:2003ch}. \ Future measurements of the double
longitudinal-spin asymmetry of charged pions, produced largely via g-q
scattering, will provide an additional handle on the magnitude of $\Delta G$%
\ and allow determination of its sign. \ 

\section{Conclusions}

RHIC, as a polarized hadron collider, provides a wealth of new opportunities
to study the spin structure of the proton. \ The accelerator has already
demonstrated success, and the RHIC community is looking forward to many more
years of running with further improvements in luminosity and polarization as
well as at higher energy. \ The PHENIX experiment has a broad program to
investigate this structure, with particular focus on the gluon's
contribution to the spin of the proton, the flavor decomposition of the sea
quarks' contributions, and the transverse spin structure of the proton. \
First results are already available, indicating that the small single
transverse-spin asymmetries seen at $x_{F}\approx 0$ at lower energies
remain small at RHIC energies and that $\Delta G$ is not large. \ The spin
structure of the proton continues to be a field of study of great interest
with much still to be explored.


%
%
%


The author wishes to express her deep appreciation to Miriam Kartch-Hughes
for making it possible to attend ISSP 2004 via a scholarship in memory of
her husband, Vernon W. Hughes. \ PHENIX acknowledges support from the
Department of Energy and NSF (U.S.A.), MEXT and JSPS (Japan), CNPq and
FAPESP (Brazil), NSFC (China), CNRS-IN2P3, CEA, and ARMINES (France), BMBF,
DAAD, and AvH (Germany), OTKA (Hungary), DAE and DST (India), ISF (Israel),
KRF and CHEP (Korea), RAS, RMAE, and RMS (Russia), VR and KAW (Sweden),
U.S.CRDF for the FSU, US-Hungarian NSF-OTKA-MTA, and US-Israel BSF.

 \bibliography{spin}


\end{document}